# Evidence of an anisotropic Griffiths phase in a two-dimensional van der Waals magnet

Anupam K. Singh[1†*], Dhananjay Nair[1*], Holger Meyerheim[1], Stuart S. P. Parkin[1†], Banabir Pal[1†]

[1]Max Planck Institute of Microstructure Physics, Halle (Saale) 06120, Germany

*Equal contribution for first author*

## Abstract

Over the last decade, the emerging field of two-dimensional van der Waals magnetism has focused mainly on long-range-ordered magnetic phases. Disordered magnetic phases, however, remain largely unexplored. Here, we report a highly anisotropic Griffiths phase driven by self-intercalation in the van der Waals magnet $Fe_5GeTe_2$. We find evidence of short-range-ordered ferromagnetic clusters embedded in a paramagnetic matrix, which dominate the magnetic response between the Curie temperature and a higher characteristic Griffiths temperature. An anomalous power-law divergence in the magnetic susceptibility, universal Lee–Yang scaling, and distinct entropy anomalies provide further evidence for this phase. Strikingly, the non-analytic Griffiths response emerges exclusively under in-plane magnetic fields, whereas out-of-plane fields abruptly alter this behavior. This directional dependence indicates that the local magnetic interactions within nanoscale clusters are closely coupled to the layered lattice geometry. Our findings establish a regime of dimensionally constrained, disorder-driven magnetism and identify self-intercalated van der Waals magnets as a platform for investigating the interplay between magnetic disorder, anisotropy, and reduced dimensionality.

† banabir.pal@mpi-halle.mpg.de, aksingh@mpi-halle.mpg.de, stuart.parkin@mpi-halle.mpg.de

## Introduction

Two-dimensional (2D) magnetism in van der Waals (vdW) materials has opened up new opportunities to explore magnetic responses in reduced dimensions[1-6]. Research over the past decade has focused primarily on 'ordered' magnetic ground states[7-9]. A multiplicity of ordered ferro- and anti-ferromagnetic phases have been shown to survive down to the monolayer limit[10-15]. On the other hand, 'disordered' magnetism, a prominent field in 3D materials[16-18], remains much less explored in the context of 2D vdW materials[19]. In particular, it remains an open question whether vdW magnets can host complex disordered magnetic states and, more importantly, how their behavior differs from that of 3D systems (Fig. 1a).

Disorder in magnetic systems arising from site dilution, off-stoichiometry and intercalation can give rise to local variations in magnetic interactions, often yielding glassy and cluster states such as spin glasses[20,21] or super-paramagnetism[22,23]. One notable example of a disorder-driven magnetic state is the Griffiths phase (GP), originally conceptualized for diluted Ising ferromagnets[24-27]. In this phase, spatially localized ferromagnetic clusters (with short-range order), embedded within a paramagnetic matrix, exhibit local magnetic order at temperatures above the Curie temperature ($T_C$), at which long-range order sets in. Between $T_C$ and a characteristic Griffiths temperature ($T_{GP}$), the magnetic response of these clusters leads to non-analytic magnetic behavior[26-28]. While the signatures of a GP are well understood in bulk materials, how the layered structure and magnetic anisotropy of vdW magnets affect a GP remains unknown. In vdW systems, these features lead to distinctly different in-plane (IP) as compared to out-of-plane (OOP) magnetic interactions, which could make Griffiths-phase effects particularly interesting. However, realizing this requires naturally occurring magnetic inhomogeneities and/or disorder driven by chemical intercalation within the vdW gaps.

Here, we report the observation of a highly anisotropic Griffiths phase driven by self-intercalation in the vdW magnet $Fe_5GeTe_2$. In particular, we find evidence that the system hosts short-range ordered (SRO) magnetic clusters that dominate the magnetic response between $T_C$ and $T_{GP}$. This is supported by anomalous power-law divergences in the susceptibility[28,29], universal Lee-Yang scaling[30-33], and distinct entropy anomalies[34]. Notably, we find that the reduced dimensionality of the vdW host renders this Griffiths phase response anisotropic: the non-analytic GP response emerges exclusively under IP magnetic fields, while the OOP fields abruptly change this behavior. This directional dichotomy indicates that the local magnetic interactions within the nanoscale magnetic clusters are closely coupled to the 2D lattice geometry, establishing a new regime of dimensionally constrained, disorder-driven magnetism.

**Crystal structure of $Fe_5GeTe_2$**

In this work, we exploit the inherent chemical inhomogeneities and the defect structure associated with self-intercalated $Fe_5GeTe_2$ (FGT5), building on prior evidence that this gives rise to magnetic inhomogeneities[35,36]. $Fe_5GeTe_2$ crystallizes in a typical layered structure in which an $Fe_5Ge$ slab is sandwiched between two Te layers, forming Te-$Fe_5$Ge-Te atomic sheets separated by the vdW gaps (Fig. 1b). To elucidate the structural disorder underlying the magnetic properties of FGT5, we have carried out a precise X-ray structural analysis using a high-brilliance liquid Ga-jet X-ray source and a six-circle diffractometer. The structure is described within the rhombohedral space group ($R3m$), with the atoms of the regular FGT5 sheets predominantly occupying the high symmetry 3a sites [for details of the atomic parameters, see Supplementary Information (SI)]. Particular attention was paid to weak additional electron density within the vdW gaps separating adjacent FGT5 sheets. The analysis used 52 symmetry-independent structure factor magnitudes obtained from integrated intensities after symmetry averaging. The best fit yields $R_U$=0.055 and a Goodness of Fit (GOF) parameter of 1.03. Using |Fobs| and the calculated phases, we reconstructed the experimental electron-density distribution, which provides a direct visualization of these structural features. Figure 1c shows the Fourier synthesis, the charge density [$\rho(x, y)$] along the c-axis [001] in the range between 0 and 0.35 lattice units. Owing to the rhombohedral symmetry, this section is sufficient to show all details of the atomic structure. As expected, the map is dominated by the large electron-density maxima associated with the two Te atoms (Te1, Te2), whereas the contributions from Fe atoms are considerably weaker. The corresponding refined structural model is shown in Fig. 1c (lower panel) and Fig. 1d in top and side view, respectively. In addition to the atoms forming the regular FGT5 sheets, the refinement reveals partially occupied Fe sites within the vdW gap.

The presence of this intercalated Fe is particularly evident from the difference-Fourier synthesis, $\Delta\rho(x, y) = \rho_{obs} - \rho_{calc}$, shown in Fig. 1e. Here, the calculated structure factors used as coefficients for the Fourier-summation were initially obtained from a model without Fe occupation of the vdW sites. After subtraction of the electron density accounted for by the nearly complete host structure, only one distinct positive residual-density maximum remains at the 3a position. The difference density is displayed on a tenfold enhanced scale, making this otherwise weak contribution readily visible. Introducing Fe at this position into the structural model accounts for the residual density and yields a site occupancy of approximately 17%. A second vdW Fe site (Wyckoff position 18c) is found from the full least-squares refinement with a considerably smaller occupancy of approximately 4%. Its contribution is correspondingly too weak to be resolved unambiguously in the difference-Fourier map. Thus, the crystallographic analysis directly establishes that the vdW gaps of the FGT5 crystal are not empty but contain a substantial amount of intrinsically disordered population of self-intercalated Fe atoms. This disorder is accompanied by incomplete occupation of several sites within the FGT5 sheets. Consequently, the crystal contains spatially distributed magnetic defects both within and between the vdW sheets, providing a natural microscopic source of variations in local magnetic exchange interactions and, hence, of the magnetic inhomogeneity enabling the Griffiths-phase behavior discussed below.

**Response of $Fe_5GeTe_2$ to in-plane magnetic field**

We first characterized the DC magnetization (M) vs temperature (T) of the system under an IP magnetic field (ab-plane, 10 Oe). Both zero-field-cooled (ZFC) and field-cooled (FC) measurements reveal a sharp magnetic phase transition near $T_C \sim 294$ K (Fig. 1f) consistent with previous reports[37,38]. However, a closer inspection of the high-temperature regime ($T > T_C$), where standard paramagnetic behavior is expected, shows two anomalous features. Firstly, we observe a step-like anomaly at around 300 K, accompanied by a subtle bifurcation between the ZFC and FC curves (left inset in Fig. 1f). Secondly, a distinct hysteresis-like opening emerges between the ZFC and FC curves near 320 K (right inset in Fig. 1f), significantly above $T_C$. Because a conventional paramagnet-to-ferromagnet transition precludes such hysteresis, these features provide the initial evidence that the system is not a simple paramagnet above $T_C$, but rather hosts persistent magnetic inhomogeneities or clusters.

To probe this deviation from conventional paramagnetism, we analyzed the temperature dependence of the inverse DC susceptibility, $\chi_{dc}^{-1}$ given in Fig. 2a at various fields, as derived from the M(T) curves. At high temperature, $\chi_{dc}^{-1}$ follows the expected Curie-Weiss (CW) linear behavior. However, as the temperature approaches $T_C$ from above, $\chi_{dc}^{-1}$ exhibits a pronounced downturn curvature evident in Fig. 2b. This downturn curvature vanishes gradually with increasing field intensity due to the significant contribution of the paramagnetic phase at higher fields (Fig.

2a). Such a downturn is characteristic of the Griffiths phase, originating from the polarization of ferromagnetic clusters of varying sizes. Each cluster possesses an effective transition temperature, generating a distribution of local critical temperatures spanning from $T_{GP}$ ~ 312 K associated with the largest clusters down to $T_C$. Notably, this downturn is progressively suppressed with increasing magnetic field, since higher fields lead to a larger contribution from the PM matrix to susceptibility.

The thermodynamic singularity associated with the GP can be classically described by a power-law divergence of the inverse susceptibility

$$\chi_{dc}^{-1} \propto (T - T_C^R)^{1-\lambda}, \quad \text{with } 0 \leq \lambda \leq 1 \qquad (1),$$

where λ is the susceptibility exponent ($0 \leq \lambda < 1$ for a GP, and $\lambda = 0$ for a paramagnet) and $T_C^R$ is the critical temperature at which the susceptibility of the largest FM clusters would diverge[28,29]. We note that $T_C^R$ is closely tied to the $T_{GP}$. Replotting this relationship on $\log_{10}$-$\log_{10}$ scale allows us to extract the critical exponents as shown in Fig. 2c. From fitting, we estimate $\lambda \approx 0.008(1)$ for the PM phase and $\lambda \approx 0.663(6)$ for the GP regions which is comparable to existing reports on complex metallic systems[39].

Besides the anomalous susceptibility, a more robust method to identify such clusters is by using a universal scaling based on the Lee-Yang Zeros model[30-33]. According to this model, the magnetization can be expressed as a function of the reduced temperature difference $t = \frac{|T-T_C|}{T_C}$ and magnetic field $h = \frac{\mu H}{k_B T_C}$ as:

$$M(t,h) = -\mu g_0 \mathrm{Im}\left[\exp\left(i\frac{A}{2h}\right) \mathrm{E}_1\left(\frac{A}{\pi} + i\frac{A}{2h}\right)\right] \qquad (2),$$

where $A = A_0\, t^{2-2\beta}$, $A_0$ and $g_0$ are constants and $\beta$ represent the critical exponent. Here μ indicates the magnetic moment of an individual atom, and $E_1$ is defined as $E_1(x) = \int_x^\infty e^{-t}/t\, dt$. Since the magnetism of FGT5 has previously been demonstrated to follow the nearly 3D Heisenberg model[40], we use $\beta = 0.35$. Using a value of $T_C \approx 294$ K, derived from the the (dM/dT) curve, we generate a theoretical scaling curve for M(T) data collected in a small field of 100 Oe (Fig. 2d). Notably, the experimental scaling curves collapse onto the theoretical scaling curve given by equation (2), confirming the formation of GP clusters in the vdW FGT5 magnet. It is worth mentioning that the presence of a GP is also independently confirmed by AC-susceptibility measurements, as discussed in the SI (see Fig. S1).

**Thermodynamic signatures of the Griffiths phase**

While the GP is defined by SRO magnetic clusters, this local ordering profoundly impacts the macroscopic magnetic entropy[34]. We extracted the isothermal magnetic entropy change ($\Delta S_M$) from isotherm M(H) data across the PM to FM transition using the Maxwell thermodynamic relation:

$$\Delta S_M = \int_{H_1}^{H_2} \left(\frac{\partial M}{\partial T}\right)_H dH \quad (3),$$

where M, H and T are magnetization, magnetic field, and temperature, respectively[34] (see Fig. S2 of the SI for more details). Because the GP survives primarily in the low-field regime, we isolated the $\Delta S_M$ response using an IP field between 20 and 300 Oe. Interestingly, the low-field $\Delta S_M$ bifurcates into two distinct peaks: a lower-temperature peak corresponding to the long-range FM transition, and a higher-temperature peak near 310 K, signifying the transition from the pure PM state to the SRO GP phase, as evident in Fig. 3a. This thermodynamic anomaly serves as an independent confirmation of the Griffiths phase in the vdW FGT5 magnet.

**Quantifying cluster dimensions via Galitski modeling**

While thermodynamic analyses confirm the presence of short-range clusters, establishing their physical dimensions is critical to linking this Griffiths phase back to the intrinsic structural defects observed in our crystallographic data (Fig. 1c-1e). To extract the physical size of these localized magnetic islands, we analyzed the DC magnetization in low field against the Galitski exponential model[41]:

$$M \propto \exp\left[-C'\left(\frac{T}{H}\right)\right] \quad (4),$$

where $C' = k_B/\mu_{avg}$ is defined by the Boltzmann constant ($k_B$) and average magnetic moment of a cluster ($\mu_{avg}$)[29]. As evidenced in Fig. 3b, the low-field data rigorously obey the Galitski model within the intermediate GP temperature window (294 K $< T <$ 312 K). Utilizing a saturated magnetization of 3.096 $\mu_B$/f.u. and the fitted value of $\mu_{avg} \sim$ 3927.2(4) $\mu_B$ derived from our 25 Oe data, we estimate the average FM cluster diameter to be ~8.3 nm. Furthermore, the presence of the SRO magnetic clusters of the GP is also verified by the modified Langevin function[29] and $\Delta S_M \sim H^2$ plot[42] analysis as described in Fig. S3.

**Anisotropic response from disordered magnetism in $Fe_5GeTe_2$:**

Having established the presence of a disorder-induced GP likely driven by self-intercalated Fe, we turn to the critical question of dimensionality: how does the intrinsic 2D layered structure of the vdW host govern this disordered state? Because the exchange interaction in vdW magnets is inherently anisotropic, with strong exchange within the vdW layer and weak interactions across the vdW gap, we anticipate a highly directional thermodynamic response. To probe this, we performed DC magnetization measurements versus temperature under an OOP magnetic field ($H \parallel$ $c$-axis, 10 Oe). As shown in Fig. 4a, the macroscopic magnetic behavior is qualitatively distinct from the response in an IP field of the same magnitude (Fig. 1f), reflecting the strong underlying magnetocrystalline anisotropy of $Fe_5GeTe_2$ (a direct comparison is given in Fig. S4). Most strikingly, we find that the anomalous step-like feature observed above $T_C$ in the IP magnetization (left inset in Fig. 1f) is completely absent under an OOP field (left inset in Fig. 4a). The second hysteretic feature also looks different (right inset in Fig. 4a). Furthermore, an analysis of the inverse susceptibility, $\chi_{dc}^{-1}$, reveals a profound directional dichotomy. Unlike the IP response, the OOP $\chi_{dc}^{-1}$ curve exhibits absolutely no downturn approaching the FM transition temperature as evident in Fig. 4b. This behavior sharply contrasts with conventional 3D disordered magnetic Griffiths phase, where nanoscale ferromagnetic clusters typically yield an isotropic GP response irrespective of the applied field direction.

To gain more insight, we also performed AC-susceptibility ($\chi_{ac}$) measurements with an OOP AC field of 10 Oe, as shown in Fig. 4c. This curve appears similar to the DC magnetization in Fig. 4a, but with a pronounced dip around 115 K, which could be attributed to a ferrimagnetic or spin-reorientation transition[37,43]. The appearance of this pronounced dip indicates that the $\chi_{ac}$ measurement is a more sensitive approach to capture any small change in magnetization ordering. Similar to the OOP DC M(T), the OOP $\chi_{ac}$ also does not exhibit any step-like feature and contains a tiny hysteresis between ZFC and FC curves above $T_C$, as depicted in inset i of Fig. 4c. Furthermore, no downturn feature appears in the $\chi_{ac}^{-1}$ vs T plot, as evident from inset ii of Fig. 4c. These features in $\chi_{ac}$ confirm the quenching of the SRO magnetic clusters of the GP under an OOP field. To understand the microscopic origin of this anisotropic quenching, one must consider both the physical geometry of the intercalated defects and the competing magnetic energies. Our high-resolution XRD analysis reveals that the self-intercalated Fe atoms reside within the vdW gaps (Fig. 1c-1e), forming localized magnetic islands that are likely quasi-two-dimensional. Consequently, the intra-cluster exchange interactions are highly robust within the ab-plane, favoring spontaneous IP ferromagnetic polarization. Moreover, the plate-like morphology of the $Fe_5GeTe_2$ crystals imposes a significant shape anisotropy (demagnetizing field), which penalizes OOP spin alignment and encourages the moments to lie within the basal plane.

## Conclusion

In summary, we find a highly anisotropic, dimensionally confined Griffiths phase in the van der Waals magnet $Fe_5GeTe_2$. Through a comprehensive magnetic and thermodynamic analysis, encompassing DC magnetization, AC susceptibility, universal Lee-Yang scaling, and magnetic entropy evolution, we demonstrate that short-range ordered ferromagnetic clusters persist well above the long-range Curie temperature. Notably, we establish that these nanoscale clusters (~8 nm) are intrinsically stabilized by structural disorder, specifically by self-intercalated Fe within the vdW gaps. Most remarkably, the reduced dimensionality of the host lattice fundamentally restricts the symmetry of this disordered state. While IP magnetic fields couple strongly to the 2D local exchange networks to reveal the non-analytic GP signatures, OOP fields abruptly quench this behavior due to strong shape anisotropy and frustrated interlayer coupling. These findings provide a microscopic framework for understanding, and potentially engineering, complex, dimensionally constrained magnetic topographies in atomically thin devices.

## Methods

Detailed magnetization measurements were performed using a MPMS3 equipped with SQUID-VSM (Quantum Design). A rectangular-shaped flat crystal is used for magnetization measurements. The possible residual field in the coil and remanence in the sample were demagnetized before the start of each measurement. Temperature-dependent magnetization M(T) data under different applied DC magnetic fields of 5-1000 Oe along the in-plane as well as out-of-plane of the samples are collected by cooling the sample from the paramagnetic (PM) to the ferromagnetic (FM) state. Note that the M(T) data are recorded under in-plane magnetic fields applied along the longest dimension of the crystal to minimize the demagnetization effects, as addressed for other systems in the literature[44]. AC-susceptibility data are also recorded with the same protocol used for M(T). Furthermore, magnetic field-dependent magnetization M(H) isotherms are recorded around the FM $T_C$ for a depth study of magnetic transition order and magnetic entropy change.

## Acknowledgments

AKS is grateful to MPI-Halle colleagues Abhay K. Srivastava, Anirban Chakraborty and Ajesh K. Gopi for informative discussions regarding sample behavior. DN is thankful to the DAAD for providing funding to visit MPI-Halle.

## References:

1 Gong, C. *et al.* Discovery of intrinsic ferromagnetism in two-dimensional van der Waals crystals. *Nature* **546**, 265-269 (2017).

2 Lee, J.-U. *et al.* Ising-Type Magnetic Ordering in Atomically Thin $FePS_3$. *Nano Lett.* **16**, 7433 (2016).

3 Burch, K. S., Mandrus, D. & Park, J.-G. Magnetism in two-dimensional van der Waals materials. *Nature* **563**, 47-52 (2018).

4 Gong, C. & Zhang, X. Two-dimensional magnetic crystals and emergent heterostructure devices. *Science* **363**, eaav4450 (2019).

5 Ningrum, V. P. *et al.* Recent advances in two-dimensional magnets: Physics and devices towards spintronic applications. *Research* (2020).

6 Park, J.-G. *et al.* 2D van der Waals magnets: from fundamental physics to applications. *Rev. Mod. Phys.* **98**, 025003 (2026).

7 Kurebayashi, H., Garcia, J. H., Khan, S., Sinova, J. & Roche, S. Magnetism, symmetry and spin transport in van der Waals layered systems. *Nat. Rev. Phys.* **4**, 150-166 (2022).

8 Papavasileiou, A. V. *et al.* Ferromagnetic Elements in Two-Dimensional Materials: 2D Magnets and Beyond. *Adv. Funct. Mater.* **34**, 2309046 (2024).

9 Zhang, B., Lu, P., Tabrizian, R., Feng, P. X.-L. & Wu, Y. 2D Magnetic heterostructures: spintronics and quantum future. *npj Spintronics* **2**, 6 (2024).

10 Huang, B. *et al.* Layer-dependent ferromagnetism in a van der Waals crystal down to the monolayer limit. *Nature* **546**, 270-273 (2017).

11 Mak, K. F., Shan, J. & Ralph, D. C. Probing and controlling magnetic states in 2D layered magnetic materials. *Nat. Rev. Phys.* **1**, 646-661 (2019).

12 Lei, S. *et al.* High mobility in a van der Waals layered antiferromagnetic metal. *Sci. Adv.* **6**, eaay6407 (2020).

13 Rahman, S., Torres, J. F., Khan, A. R. & Lu, Y. Recent developments in van der Waals antiferromagnetic 2D materials: Synthesis, characterization, and device implementation. *ACS Nano* **15**, 17175-17213 (2021).

14 Ziebel, M. E. *et al.* CrSBr: an air-stable, two-dimensional magnetic semiconductor. *Nano Lett.* **24**, 4319-4329 (2024).

15 Cheon, C.-Y. *et al.* Nature of 2D XY antiferromagnetism in a van der Waals monolayer. *Nat. Commun.* (2025).

16 Silevitch, D. *et al.* A ferromagnet in a continuously tunable random field. *Nature* **448**, 567-570 (2007).

17 Karube, K. *et al.* Disordered skyrmion phase stabilized by magnetic frustration in a chiral magnet. *Sci. Adv.* **4**, eaar7043 (2018).
18 Armstrong, S., Silevitch, D. & Rosenbaum, T. Experimental consequences of disorder at an antiferromagnetic quantum phase transition. *Nat. Commun.* **16**, 2178 (2025).
19 Pal, B. *et al.* Realization of a Spin Glass in a two-dimensional van der Waals material. *Science* **392**, 624-628 (2026).
20 Binder, K. & Young, A. P. Spin glasses: Experimental facts, theoretical concepts, and open questions. *Rev. Mod. Phys.* **58**, 801 (1986).
21 Goremychkin, E. *et al.* Spin-glass order induced by dynamic frustration. *Nat. Phys.* **4**, 766-770 (2008).
22 Bean, C. & Livingston, u. D. Superparamagnetism. *J. Appl. Phys.* **30**, S120-S129 (1959).
23 Lachman, E. O. *et al.* Visualization of superparamagnetic dynamics in magnetic topological insulators. *Sci. Adv.* **1**, e1500740 (2015).
24 Griffiths, R. B. Nonanalytic behavior above the critical point in a random Ising ferromagnet. *Phys. Rev. Lett.* **23**, 17 (1969).
25 Bray, A. Nature of the Griffiths phase. *Phys. Rev. Lett.* **59**, 586 (1987).
26 Lu, R. *et al.* Evidence for a Griffiths Phase to Cluster Spin Glass Transition in the $La_{2/3}Sr_{1/3}(Mn_{1-3x}Al_{2x}Ti_{x})O_3$ System. *Adv. Sci.* **11**, 2408517 (2024).
27 Salamon, M. B., Lin, P. & Chun, S. H. Colossal magnetoresistance is a Griffiths singularity. *Phys. Rev. Lett.* **88**, 197203 (2002).
28 Jiang, W., Zhou, X., Williams, G., Mukovskii, Y. & Glazyrin, K. Is a Griffiths phase a prerequisite for colossal magnetoresistance? *Phys. Rev. Lett.* **99**, 177203 (2007).
29 Saha, S., Dutta, A., Gupta, S., Bandyopadhyay, S. & Das, I. Origin of the Griffiths phase and correlation with the magnetic phase transition in the nanocrystalline manganite $La_{0.4}(Ca_{0.5}Sr_{0.5})_{0.6}MnO_3$. *Phys. Rev. B* **105**, 214407 (2022).
30 Blythe, R. & Evans, M. Lee-Yang zeros and phase transitions in nonequilibrium steady states. *Phys. Rev. Lett.* **89**, 080601 (2002).
31 Chan, P. Y., Goldenfeld, N. & Salamon, M. Critical behavior of Griffiths ferromagnets. *Phys. Rev. Lett.* **97**, 137201 (2006).
32 Peng, X. *et al.* Experimental observation of Lee-Yang zeros. *Phys. Rev. Lett.* **114**, 010601 (2015).
33 Ovchinnikov, A. *et al.* Unbiased identification of the Griffiths phase in intercalated transition metal dichalcogenides by using Lee-Yang zeros. *Phys. Rev. B* **106**, L020401 (2022).
34 Ghosh, K., Mazumdar, C., Ranganathan, R. & Mukherjee, S. Griffiths phase behaviour in a frustrated antiferromagnetic intermetallic compound. *Sci. Rep.* **5**, 15801 (2015).
35 Gopi, A. K. *et al.* Thickness-tunable zoology of magnetic spin textures observed in $Fe_5GeTe_2$. *ACS Nano* **18**, 5335 (2024).
36 Silinskas, M. *et al.* Self-Intercalation as Origin of High-Temperature Ferromagnetism in Epitaxially Grown $Fe_5GeTe_2$ Thin Films. *Phys. Rev. Lett.* **133**, 256702 (2024).
37 Ly, T. T. *et al.* Direct observation of Fe-Ge ordering in $Fe_{5-x}GeTe_2$ crystals and resultant helimagnetism. *Adv. Funct. Mater.* **31**, 2009758 (2021).
38 Deng, Y. *et al.* Layer-number-dependent magnetism and anomalous Hall effect in van der Waals ferromagnet $Fe_5GeTe_2$. *Nano Lett.* **22**, 9839-9846 (2022).
39 Tian, F. *et al.* Griffiths phase arising from local lattice distortion and spin glass above the Curie temperature in $Ni_2MnSb$ polycrystalline Heusler alloy. *Phys. Rev. B* **109**, 224405 (2024).
40 Li, Z. *et al.* Magnetic critical behavior of the van der Waals $Fe_5GeTe_2$ crystal with near room temperature ferromagnetism. *Sci. Rep.* **10**, 15345 (2020).
41 Galitski, V., Kaminski, A. & Das Sarma, S. Griffiths phase in diluted magnetic semiconductors. *Phys. Rev. Lett.* **92**, 177203 (2004).
42 Pakhira, S., Mazumdar, C., Ranganathan, R., Giri, S. & Avdeev, M. Large magnetic cooling power involving frustrated antiferromagnetic spin-glass state in $R_2NiSi_3$ (R= Gd, Er). *Phys. Rev. B* **94**, 104414 (2016).

43 May, A. F. *et al*. Ferromagnetism near room temperature in the cleavable van der Waals crystal $Fe_5GeTe_2$. *ACS Nano* **13**, 4436-4442 (2019).

44 Jiang, W., Zhou, X., Williams, G., Mukovskii, Y. & Glazyrin, K. Griffiths phase and critical behavior in single-crystal $La_{0.7}Ba_{0.3}MnO_3$: Phase diagram for $La_{1-x}Ba_xMnO_3$ ($x \leqslant 0.33$). *Phys. Rev. B—Condens. Matter Mater. Phys.* **77**, 064424 (2008).

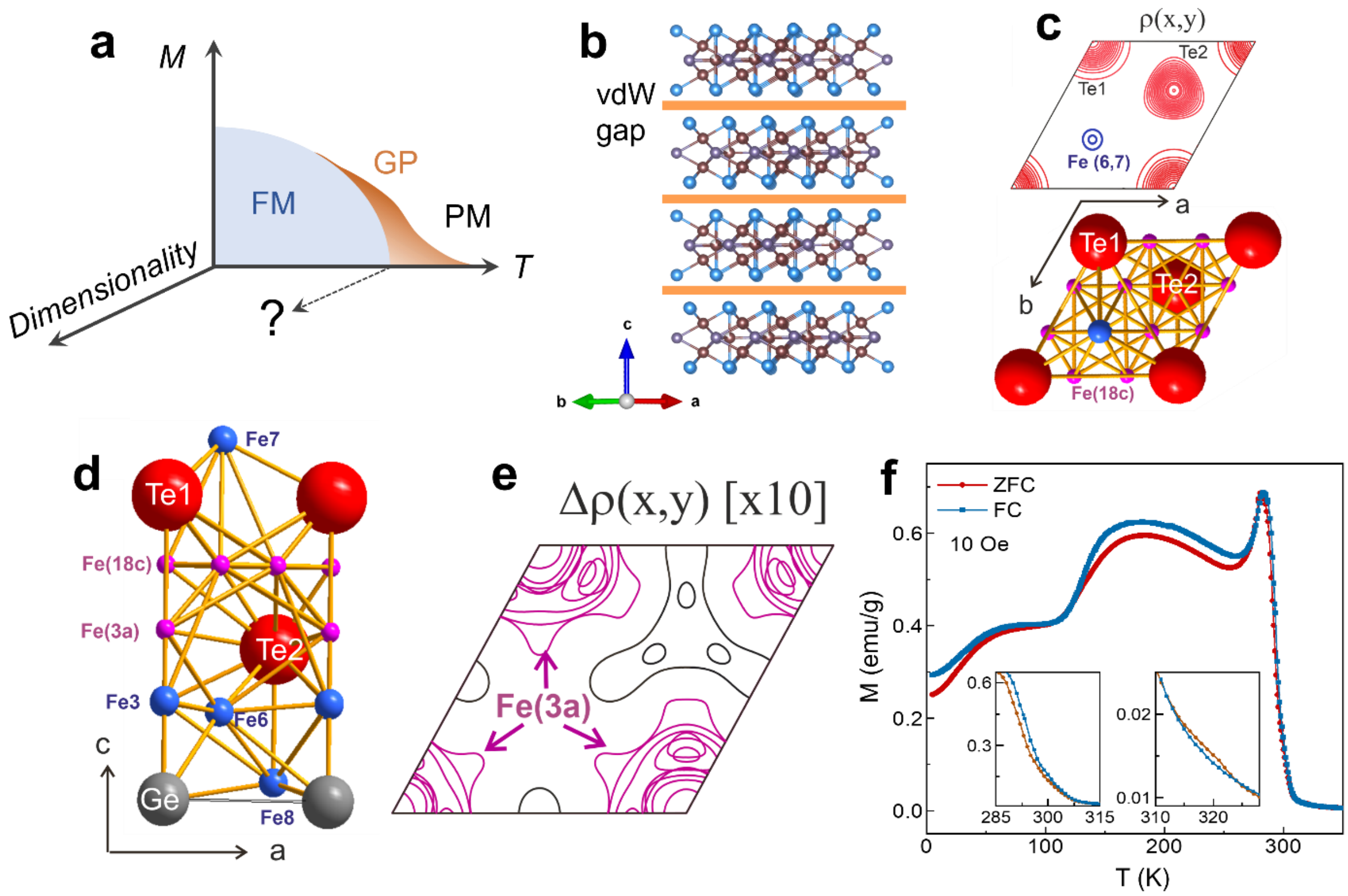

**Fig. 1**: **Schematic of disorder-induced GP, structural and magnetic characterizations of an FGT5 crystal. a** Schematic showing the disorder-induced Griffith phase space. **b** Crystal structure of FGT5 where different-colored balls represent Fe (brown), Ge (purple) and Te (blue). FGT sheets are separated by vdW gaps highlighted by the orange lines where the presence of disorder can give rise to short-range magnetic clusters of the GP. **c** Top: Calculated charge density ρ(x,y) projected along [001] between z = 0 and z = 0.35. Bottom: Top view of the corresponding structural model, while the **s**ide view is given in **d**. ρ(x,y) is dominated by the Te atoms #1 and #2. **e** Difference charge density Δρ(x, y) calculated by neglecting vdW site atoms Fe(18c) and Fe(3a). The latter directly appear in Δρ(x, y). **f** Temperature-dependent magnetization M(T) measured under an in-plane magnetic field of 10 Oe during zero field-cooled (ZFC) and field cooling (FC) cycles. Insets show enlarged views of M(T) around the ferromagnetic transition.

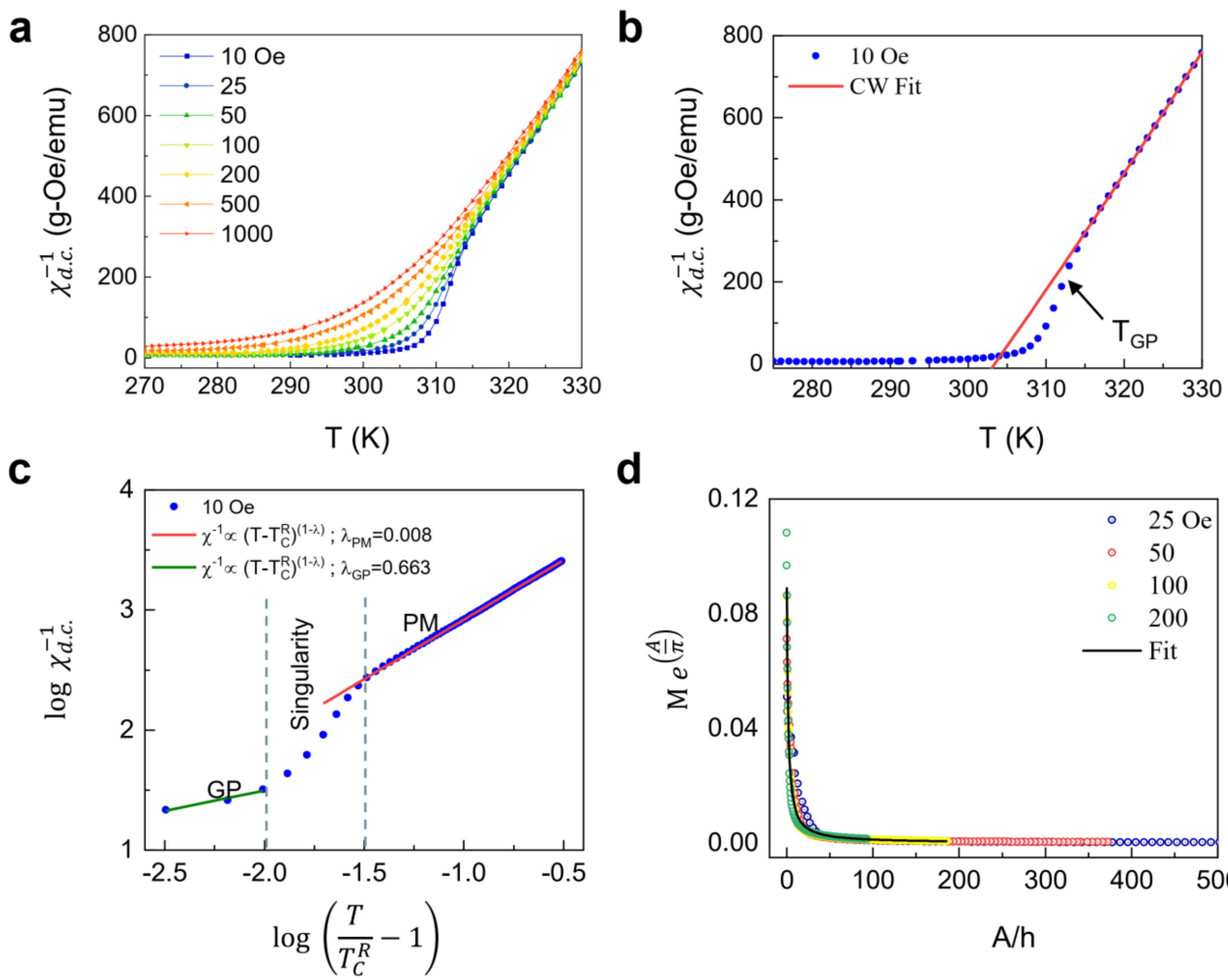

**Fig. 2**: **Analysis of magnetization data collected under IP field. a** Inverse magnetic susceptibility ($\chi_{dc}^{-1}$) vs T obtained from M(T) collected at the indicated fields. **b** Curie-Weiss (CW) fitting of $\chi_{dc}^{-1}$ vs T data at 10 Oe, where the Griffiths temperature is marked by $T_{GP}$. **c** Modelling of $\chi_{dc}^{-1}$ using a power law (eq. 1) at 10 Oe plotted on a $\log_{10}$-$\log_{10}$ scale and indicated by solid lines. The GP and paramagnetic phase (PM) are separated by a singularity region marked by dashed lines. **d** Lee-Yang Zeros analysis where $Me^{A/\pi}$ vs A/h is plotted at the indicated fields. The solid black curve indicates the theoretical model using a universal scaling function (eq. 2).

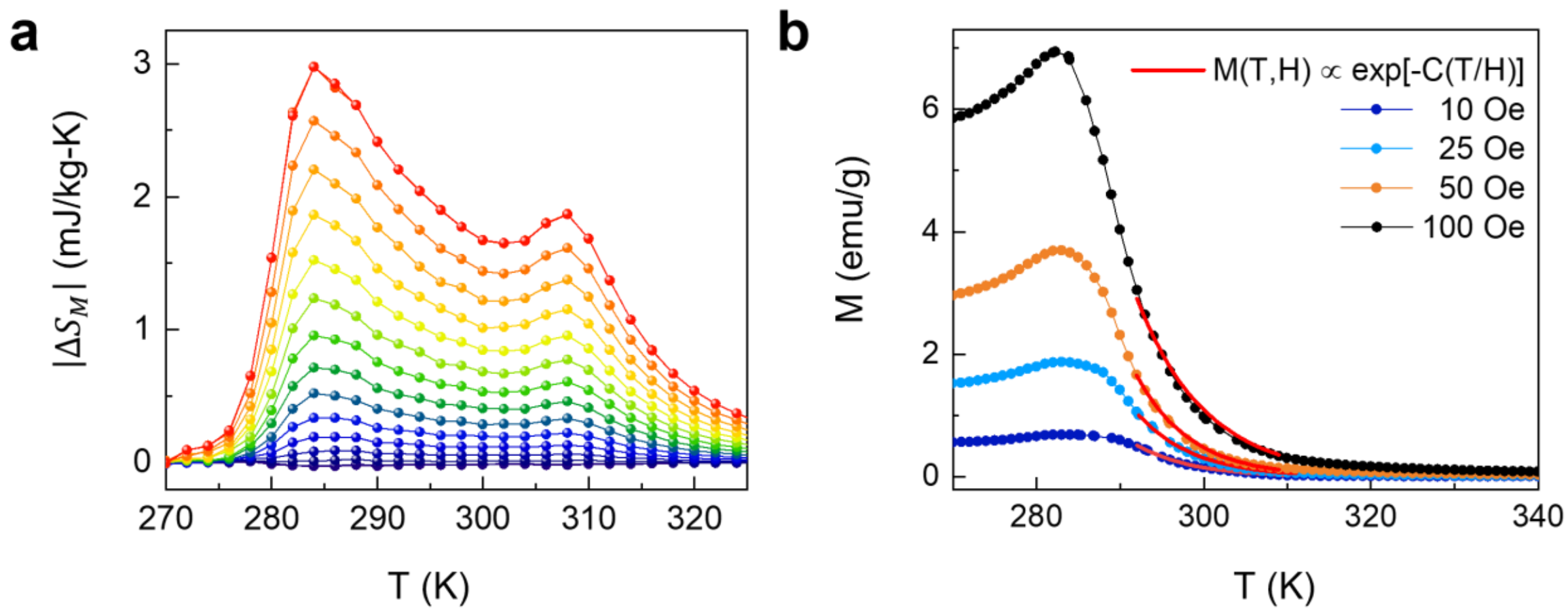

**Fig. 3**: **Analysis using magnetic entropy and the Galitski model. a** Isothermal magnetic entropy change ($\Delta S_M$) vs T plot in the low-field range (20 Oe to 300 Oe), obtained by applying the Maxwell thermodynamic relation (eq. 3) to the measured isotherms. The appearance of two peaks is characteristic of FM and GP transitions, respectively. **b** Modelling of IP M(T) data using the Galitski model given in eq. (4). The fit is shown by solid red curves at the indicated fields.

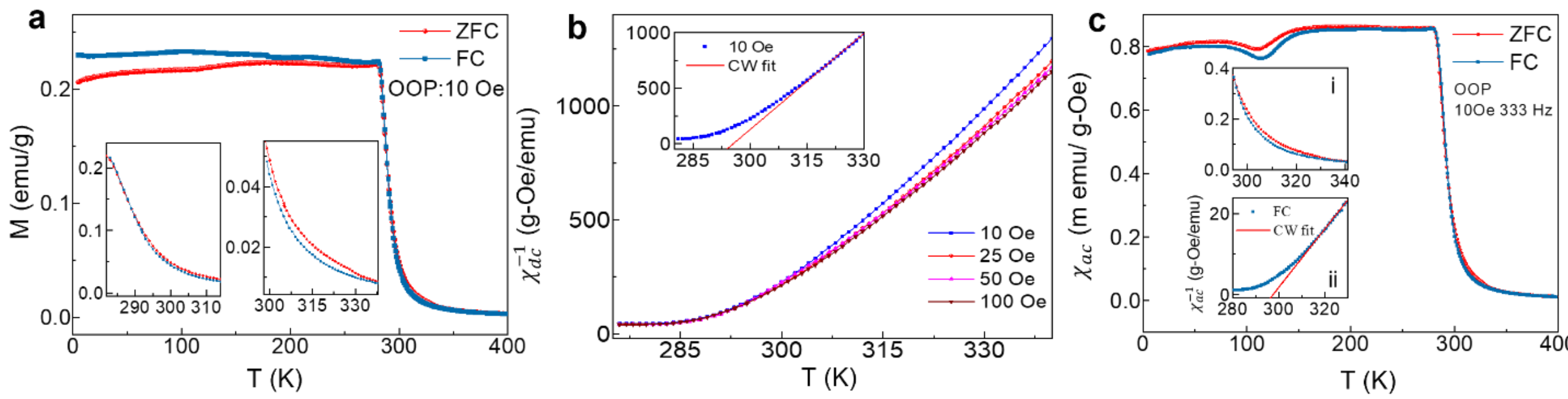

**Fig. 4**: **Out-of-plane (OOP) magnetic properties**. **a** Temperature-dependent magnetization M(T) measured under an OOP magnetic field of 10 Oe during zero field-cooled (ZFC) and field cooling (FC) cycles. Insets show an enlarged view of M(T) around the FM transition. **b** Inverse magnetic susceptibility ($\chi_{dc}^{-1}$) vs T plot obtained from the M(T) collected at the indicated fields. Inset of **b** shows Curie-Weiss (CW) fitting using the $\chi_{dc}^{-1}$ vs T plot at 10 Oe. **c** Temperature-dependent AC-susceptibility ($\chi_{ac}$) measured under an OOP AC field amplitude of 10 Oe at a frequency of 333 Hz during ZFC and FC cycles. Insets i shows an enlarged view of $\chi_{ac}$ around the FM transition and inset ii shows CW fitting of the $\chi_{ac}^{-1}$ vs T plot.